# Deposition of highly-crystalline AlScN thin films using synchronized HiPIMS – from combinatorial screening to piezoelectric devices

Jyotish Patidar[1], Kerstin Thorwarth[1], Thorsten Schmitz-Kempen[2], Roland Kessels[2], Sebastian Siol[1*]

[1] Empa, Swiss Federal Laboratories for Materials Science and Technology, Dübendorf, Switzerland

[2] aixACCT Systems GmbH, Aachen, Germany

*Corresponding author:*

*Sebastian Siol, Sebastian.Siol@empa.ch*

## TOC Figure

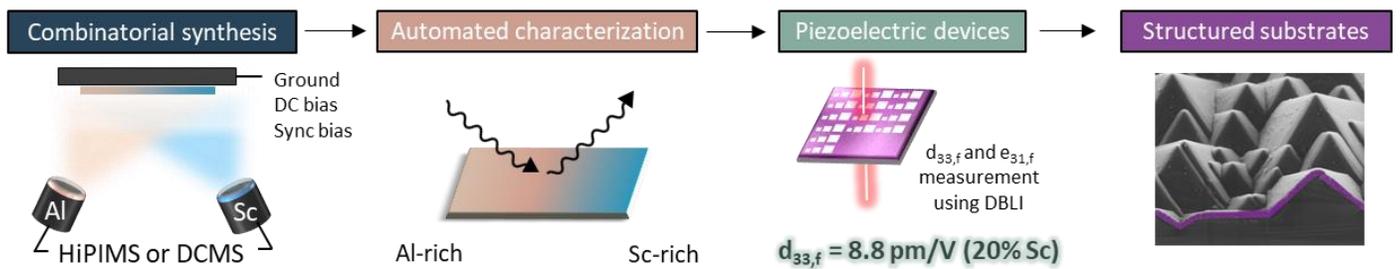

## Abstract

Fueled by the 5G revolution, the demand for advanced radio frequency micro-electromechanical systems (MEMS) based on AlScN is growing rapidly. However, synthesizing high-quality, textured AlScN thin films is challenging. Current approaches typically rely on high temperatures and expensive compound targets. In this study, we demonstrate the feasibility of ionized physical vapor deposition to deposit highly oriented AlScN films with minimal defects at lower temperatures. Using metal-ion synchronized high-power impulse magnetron co-sputtering (MIS-HiPIMS) we can selectively bombard the growing film with Al and/or Sc ions to enhance the adatom mobility while simultaneously providing the ability to tune stress and coat complex structures conformally. We find that the Sc solubility in wurtzite AlN is slightly reduced, whereas crystallinity and texture are markedly improved. Disoriented grains, a key challenge in growing AlScN films, are completely removed *via* substrate biasing, while the residual stress can be tailored by adjusting the same. The measured piezoelectric response of the films is in line with DFT predictions and on par with the current state of the art. Finally, we demonstrate conformal deposition of c-axis textured AlScN on structured Si wafers underlining the promise of ionized PVD for the fabrication of advanced RF filters and next-generation MEMS devices.

# 1. Introduction

Fueled by recent advances in wireless communication our interaction with technology has undergone a profound transformation. The higher frequencies and wider bandwidths associated with the current 5G and the forthcoming 6G standards necessitate the development of high-performance radio frequency (RF) filters. While these RF filters are predominantly used in the telecommunication industry, their applications extend to various emerging technologies, including unmanned drones, satellites, self-driving vehicles, wearables, and Internet of Things (IoT) applications. State-of-the-art filter designs are largely based on piezoelectric micro-electromechanical systems (MEMS). In RF filters based on MEMS acoustic resonators, the frequency and bandwidth of the filter critically depend on the properties of the underlying thin-film piezoelectric material. Aluminum Nitride (AlN) in wurtzite structure has long been recognized as a valuable material in RF filters due to its linear frequency response and CMOS-compatible processing, however, it suffers from a comparably low piezoelectric response with low electromechanical coupling.[1], [2] The addition of Scandium (Sc) to the AlN lattice has been shown to introduce unique advantages, such as improved piezoelectric coefficients and enhanced electromechanical properties. [3], [4] With their improved functional properties and the ability to tailor their characteristics through compositional and stress engineering, AlScN films have emerged as the material of choice for 5G RF-components and are paving the way for exciting advancements in the field of piezoelectric MEMS[5]–[7].

The synthesis of Aluminum Scandium Nitride (AlScN) films presents several challenges that need to be overcome to extract their full potential in piezoelectric applications. For piezoelectric applications, crystalline quality and texture are critical properties which affect device performance. To ensure a high average piezoelectric response, the films need to exhibit uniform c-axis texture and polarization. However, increasing the amount of Sc in the AlN lattice introduces structural frustration causing local distortions, structural defects, and the growth of abnormally oriented grains.[8]–[10] Using high deposition temperatures can help in mitigating these defects, however, due to the high mixing enthalpy of wurtzite-AlN and cubic-ScN, this also promotes the precipitation of secondary phases. [11], [12] In addition, the use of higher process temperatures on an industrial scale leads to higher production costs and reduced sustainability. Residual stress is another critical factor affecting the performance of RF-MEMS devices. In particular, stress can negatively influence the frequency response and mechanical integrity, emphasizing the importance of stress-free, crystalline, and uniformly textured AlScN films for commercial viability and device integration.[13] Co-optimizing these properties poses significant challenges for the process development.

Apart from the challenging synthesis of crystalline AlScN films with the aforementioned required properties, novel technologies in MEMS devices also require uniform coatings on parts with complex topographies such as cantilevers, resonators, micro-trenches, 3-D electromechanical metamaterials, and so on.[14], [15] Conformal deposition in trench-like structures with texture along the substrate normal is of great interest in the MEMS industry right now to increase the active surface area of the piezoelectric materials. Atomic layer deposition (ALD) and metalorganic chemical vapor deposition (MOCVD) are usually explored for such applications due to their ability to coat complex structures.[16], [17] However, ALD processes typically feature low adatom kinetic energy and deposition rates, while MOCVD requires high growth temperatures, leading to high production costs. Thus, finding a novel deposition route that could provide conformal coatings irrespective of deposition angles and geometries, and could surpass the drawbacks of ALD and MOCVD would unlock a number of exciting new applications such as piezoelectric sensors or actuators on 3D printed parts with complex shapes or on biomedical devices and wearables.

These challenges motivate the development of advanced deposition techniques, which enable the careful control of the growth parameters to achieve high-quality AlScN films with tailored structural and functional properties. The most common approaches to date to manufacture AlScN thin films at an industrial scale are direct current magnetron sputtering (DCMS) and pulsed direct current magnetron sputtering (pDCMS) from compound AlSc targets.[18]–[20] To unlock the full potential of AlScN films for future piezoelectric and ferroelectric applications researchers are continuously exploring alternative synthesis approaches and material engineering strategies.[18], [21]–[26] In this regard, the development of Ionized PVD (IPVD) methods, such as high power impulse magnetron sputtering (HiPIMS), and their use in the synthesis of functional films have picked up interest in the last few years.[10], [27], [28] Due to the inherently very high ionization rates of the process, the film-forming ions can be accelerated onto the growing film using substrate bias potentials, which helps in attaining high adatom mobility to deposit crystalline films at lower temperatures while simultaneously enabling conformal coverage over structured surfaces.[29], [30] [31]–[34] Despite these advantages, until now, HiPIMS deposition approaches have been predominantly used for hard coatings, optical coatings or metallization.[34]–[37] The main reason for this is, that thin films deposited by HiPIMS often exhibit defects and large residual stresses caused by process gas incorporation (typically $Ar^+$ ions) which limits their feasibility for defect-sensitive applications. These effects can be mitigated by the application of a novel metal-ion-synchronized HiPIMS (MIS-HiPIMS) approach, which is based on the selective acceleration of the film-forming metal-ions by applying a pulsed substrate bias synchronized to the metal-rich part of the plasma.[28], [38] We recently demonstrated the deposition of out-of-plane (0002) textured AlN films using metal-ion synchronized HiPIMS. By accelerating only the Al-ions with low to moderate substrate bias potentials remarkable improvements in the films' crystallinity and texture were achieved. The dense microstructure also resulted in a significantly improved resistance against bulk oxidation. Most importantly though, we could demonstrate that the new process design enables the deposition of highly-textured films on structured surfaces and with shallow deposition angles.[28] These results highlight the potential of IPVD processes for the deposition of functional thin films, especially when highly textured and crystalline films are required. Consequently, such a deposition principle should be ideally suited to deposit high-quality AlScN thin films. To our knowledge, no synchronized HiPIMS deposition for AlScN has been reported to date.

The additional degrees of freedom during the reactive co-sputtering of AlScN using IPVD techniques provide more flexibility for tuning the film's properties but also pose challenges in process optimization. Besides the conventional sputter process parameters, such as working pressure, nitrogen partial pressure, or sputter fluxes, reactive IPVD co-sputtering with substrate bias synchronization opens up several new dimensions in the process parameter space. This necessitates a meticulous and mindful control of sputter conditions, which can greatly influence the important film properties including structure, phase-purity, and residual stress.

Of particular importance for AlScN thin film development is the non-equilibrium Sc solubility in AlN, since higher Sc-content has been shown to not only increase the piezoelectric response but also lower the coercive field for ferroelectric switching.[39] Despite its importance, only a few studies to date report on the non-equilibrium Sc solubility as a function of the synthesis environment. This is understandable, as traditional synthesis approaches require the deposition of many samples with varying compositions along with their individual structural analysis. A promising route to this challenge lies in the application of high-throughput experiments such as combinatorial screening.[12], [40] Using gradient deposition, large areas of the synthesis phase space can be covered on a single materials library. Coupling this deposition approach with automated characterization and data analysis facilitates the rapid investigation of the phase-constitution (i.e. the precipitation of secondary phases) and crystallinity as a function of composition for a given set of synthesis parameters.

In this work, we make use of such techniques to investigate the feasibility of different IPVD approaches to deposit highly-textured and crystalline AlScN films. The workflow of this investigation is shown in **Figure 1**. Co-sputtering from Al and Sc targets in a reactive atmosphere is performed using a hybrid deposition which employs different combinations of DCMS and/or HiPIMS. In addition, different biasing strategies are developed to probe the effect of low-energy ion bombardment on the crystalline quality and stress-state of the films. A comparison of the crystalline quality and phase-purity for a range of Sc compositions is performed based on an extensive combinatorial screening. Subsequently, based on the observed crystallinity and the Sc-solubility limits the most promising deposition schemes are chosen. Finally, for each approach optimized singe-phase films are deposited with approximately 20 at.% Sc. These films are characterized for their crystallinity, residual stress, surface morphology, and most importantly their piezoelectric coefficients ($d_{33,f}$).

The results of this study clearly show that the added ad-atom mobility of the IPVD approaches does not only lead to highly oriented and crystalline AlScN thin films at moderate substrate temperatures. Moreover, the biasing of the substrate provides additional advantages like the removal of disoriented grains and the ability to tune stress by adjusting the kinetic energy of the film-forming species. In addition, despite the shallow deposition angle, the films show piezoelectric coefficients close to the current state-of-the-art. At the same time, the novel deposition approach offers the promise of lower deposition temperatures and uniform deposition on structured surfaces which could pave the way to exciting new applications and advanced RF-MEMS devices in the future.

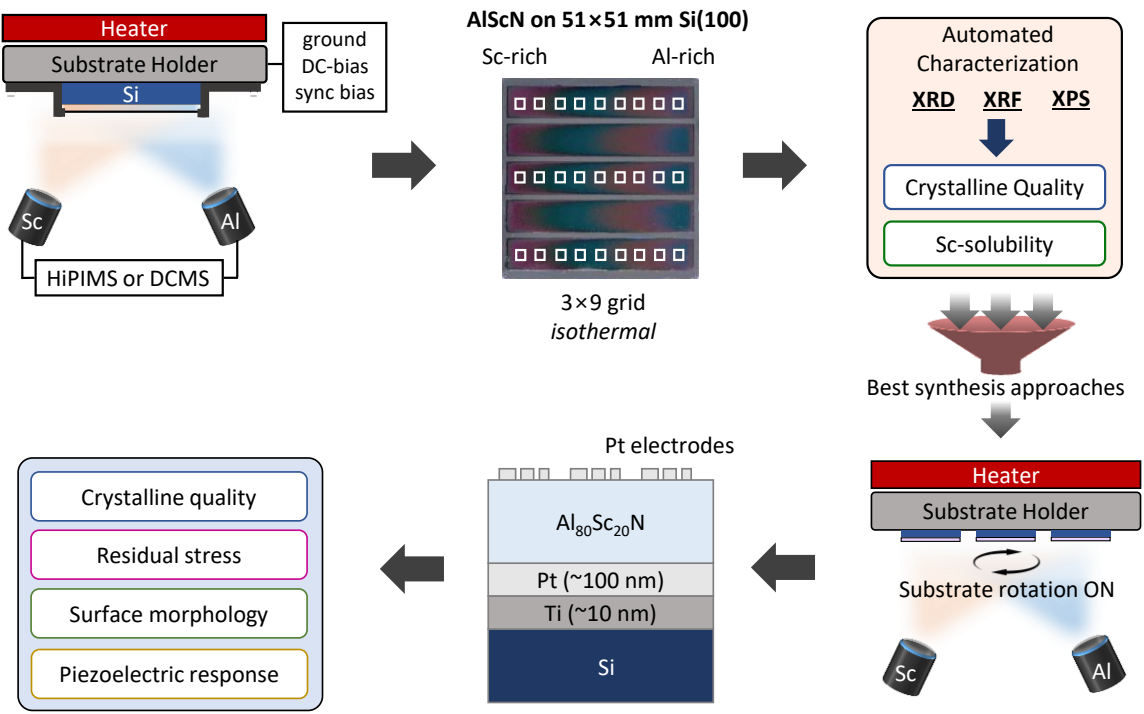

**Figure 1:** Workflow used in this study. Combinatorial materials libraries are deposited on Si substrates for different combinations of sputtering modes and substrate biasing. The libraries are characterized using XRD-, XRF-, and XPS-mapping to determine the relative crystalline quality and Sc-solubility. Based on the most promising synthesis schemes from the combinatorial screening single-phase $Al_{80}Sc_{20}N$ films are synthesized and piezoelectric devices are fabricated with Pt top and bottom electrodes.

## 2. Results

### 2.1. Synchronization of the Substrate-Bias Potential

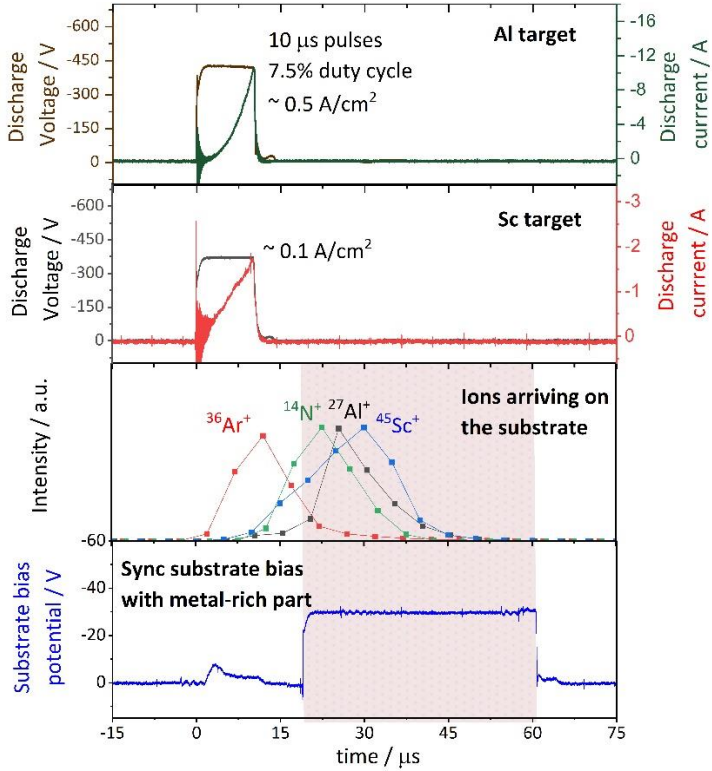

**Figure 2:** Synchronization of substrate bias pulse with the metal-rich part of HiPIMS plasmas. The pulses on Al and Sc targets are also synchronized to start at the same time.

During HiPIMS, following each sputter pulse, the different ionic species arrive at the substrate at different times. This is due to process-gas rarefaction and differences in ionic mass.[41] In metal-ion synchronized HiPIMS this phenomenon is leveraged by synchronizing a negative substrate bias potential to selectively accelerate the film-forming metal-ions, rather than the process gas ions. The goal is to deposit high-quality thin films at low deposition temperatures, while simultaneously minimizing process gas incorporation. For an effective synchronization, the time-of-flight of the Al and Sc ions has to be known.[28] To this end, we performed time- and energy-resolved mass spectrometry using a Hiden Analytical EQP-300 placed at the working distance while facing the respective sputter guns. The measurements were performed for each ion species, including $^{14}$N, $^{36}$Ar, $^{27}$Al, and $^{45}$Sc. A less abundant isotope of Ar is used here to avoid the saturation of the detector. The TOF-measurements along with the I-V curves of the HiPIMS pulses are shown in **Figure 2**.

In all experiments, the Sc and Al HiPIMS pulses start at the same time (i.e. 0 µs). It is apparent from the TOF measurements that the Ar-ions arrive predominantly in the first 15 µs after the initiation of the HiPIMS pulses, whereas the Al and Sc ion fluxes reach peak intensity at X µs and Y µs, respectively. We find that applying a 40 µs substrate bias pulse with a time delay of 20 µs effectively accelerates both Al and Sc ions while avoiding acceleration of the process gas ions. A moderate potential of -30 V is chosen to achieve a median ion kinetic energy below the lattice displacement threshold.[28] The motivation behind this approach is to provide a low-energy metal-ion bombardment while simultaneously avoiding ion implantation, which could lead to structural defects and excess compressive stress.

## 2.2. Combinatorial Screening of Deposition Modes

The reactive co-sputtering of AlScN with synchronized ionized PVD processes poses significant challenges to process development due to the vast synthesis parameter space. In this study, we try to investigate the feasibility of different principle deposition modes for the deposition of high-quality AlScN thin films with the help of combinatorial high-throughput experiments. For the films' deposition, the Al and Sc sputter guns were operated in multiple combinations of sputtering modes, namely HiPIMS-AlN/DCMS-ScN, DCMS-AlN/HiPIMS-ScN, DCMS-AlN/DCMS-ScN, HiPIMS-AlN/HiPIMS-ScN. These modes correspond to ionizing different fractions of the sputter flux, i.e. Al, Sc, or both Al and Sc. For the substrate, different biasing strategies are evaluated. These include a grounded substrate (no additional acceleration), a constant bias of -30 V (a constant acceleration of all ions, including $Ar^+$), or a synchronized bias pulse of -30 V (selective metal-ion acceleration).

For each combination of deposition modes several combinatorial materials libraries were deposited covering a wide range of Sc alloying concentrations. In total over 200 unique samples were tested. The samples were evaluated for crystalline quality and phase constitution through automated XRD analysis. The Full Width at Half Maximum (FWHM) of the (002) AlScN peak serves as an indicator of grain size and, consequently, is utilized as a measure of the films' crystalline quality. **Figure 3** illustrates the FWHM values for all examined samples in the combinatorial libraries, organized based on decreasing solubility limits from (a) to (g). The solubility limit is calculated by evaluating the phase purity of w-AlScN films through XRD θ-2θ screening based on the emergence of the c-ScN phase with varying Sc concentration. It is well known that for DCMS depositions at oblique angles, the films preferentially grow in the direction of the sputter source, according to the tangent rule.[42], [43] This phenomenon was evident in the combinatorial screening, where the flux from the sputter gun connected to DC power supplies promoted the growth of the respective phase in the direction of the deposition flux. Thus, it is critical to look for secondary phases with the help of χ scans to avoid an overestimation of the phase purity and solubility limits. The procedure for determining the solubility limits is illustrated in detail in **Supporting Information S1**.

The solubility limit is marked by a red dotted line in the **Figure 3**. Pronounced precipitation typically leads to higher leakage currents and reduced crystallinity, which can negatively affect device performance. A more direct correlation exists between the structural properties and the piezoelectric response in AlScN. For this reason, the region below FWHM of 0.5° is marked green and labelled as the region of interest. The range of interest (FWHM from 0 to 0.5°) chosen is relatively generous here since the grain size of the films will certainly increase with substrate rotation and optimized synthesis parameters for single-phase samples. As anticipated, the FWHM of AlScN notably increases beyond the solubility limit for all deposition modes. Strikingly, some deposition modes, (see **Figure 3 (e-g)**) seem less susceptible to this degradation. This is likely due to the increased ion bombardment, particularly when applying substrate biasing, which leads to higher adatom mobility, which helps in keeping the structural integrity intact even after slight precipitation. Films deposited using DC on both sputter guns exhibited markedly higher FWHM due to lower ionized flux and the inherent low adatom mobility. The introduction of ion bombardment, facilitated by HiPIMS on either or both sputter guns, demonstrated improved crystalline properties irrespective of substrate biasing. The percentage of ion flux in HiPIMS mode for each sputter gun was calculated for both Al and Sc by depositing the films with a grounded substrate and with a positive bias. The ion-to-neutral flux was then calculated by taking the ratio of the difference in thickness of the film with and without bias. The resulting ion flux for Al and Sc in HiPIMS mode at 0.5 and 0.1 $A/cm^2$ was found to be approximately 30% and 9%, respectively. The ionization fraction of Scandium (Sc) plasma is relatively lower, primarily due to the constraints imposed by the sputter power limitations to achieve lower alloying percentages. Although Sc in HiPIMS modes provides a lower ion-flux density, the ion mass and therefore the momentum transfer to the growing film is higher when compared to Al. Consequently, no significant difference in film quality was observed when comparing selective Al- or Sc-ion irradiation of the growing film. A more pronounced difference was observed for the different substrate biasing modes. In particular films with Al in HiPIMS mode and Sc in DCMS mode showed an increase in crystallinity with the application of a substrate bias. The acceleration of Sc ions appears to be detrimental which could be due to the higher mass and consequently higher momentum transfer to the growing film. In addition to changes in grain size a lower Sc solubility limit was observed with the application of substrate biasing. Here films with DC substrate bias exhibited the lowest Sc solubility.

Overall, the grain size of HiPIMS deposited films was far superior compared to the DCMS deposition (**Figure 3(a)**). In addition, the AlScN grains orientation varied across the library in the direction of sputter flux, as confirmed by XRD. This is likely a result of the shallow deposition angles in our confocal sputter chamber geometry combined with the moderate substrate temperature used in this study.[28] These results underline that for best results during DCMS of AlScN thin films on-axis deposition at higher deposition temperatures is necessary.[44], [45] Considering these factors, we focus on the following detailed characterization of the films deposited with IPVD approaches.

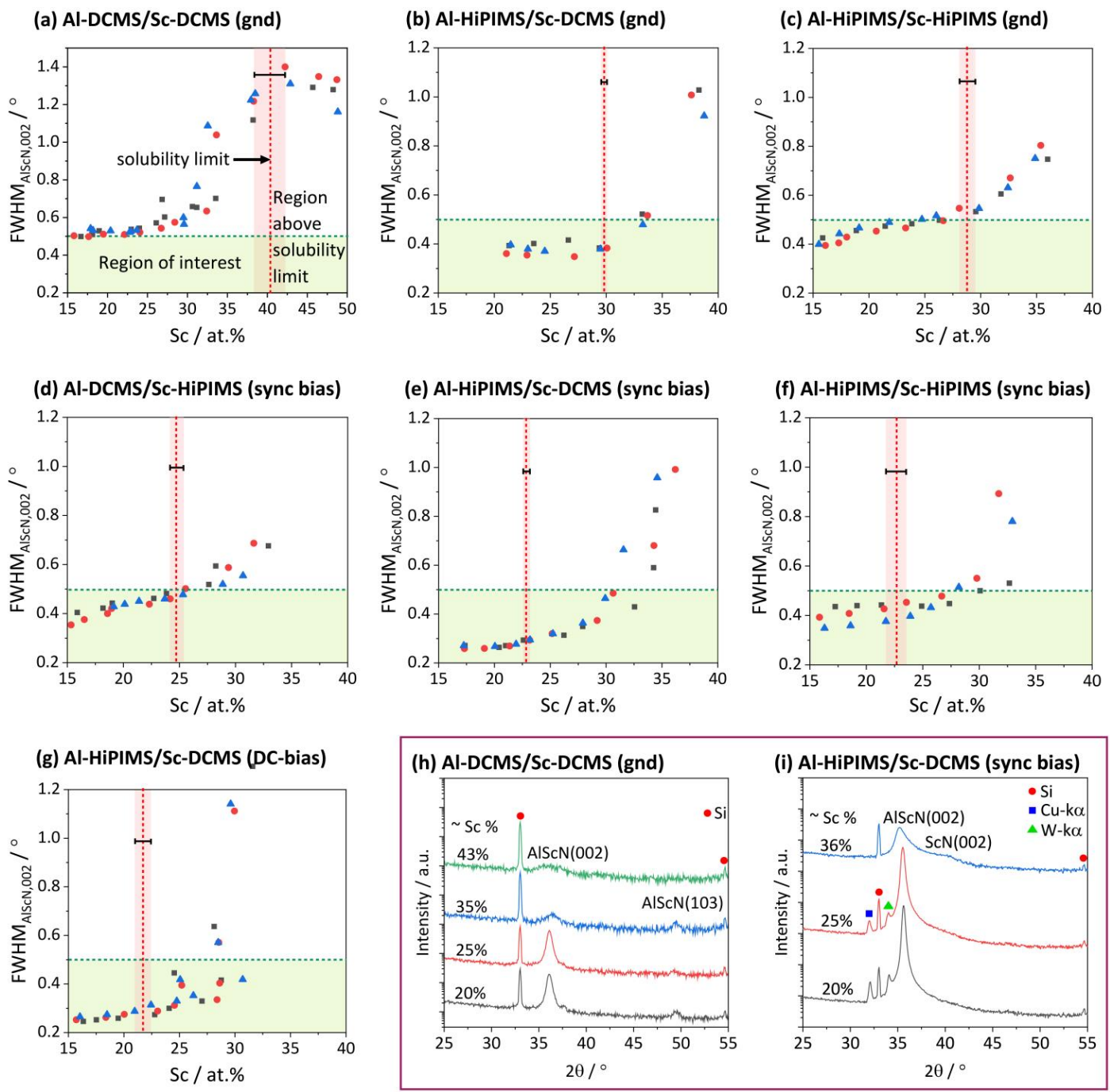

**Figure 3.** Relative comparison of crystalline quality and estimation of solubility limits of AlScN combinatorial libraries. The combinatorial libraries are deposited for different combinations of sputter modes and substrate biasing approaches (labelled in respective plots (a-g)). The FWHM of the AlScN (002) reflection is used as a proxy for the grain size. The red line marks the first detectable precipitation of c-ScN. To illustrate the formation of precipitates, the XRD patterns of films deposited with Al-DCMS/Sc-DCMS (ground) and Al-HiPIMS/Sc-DCMS (sync bias) are stacked for different Sc composition in Figure (h) and (i), respectively.

## 2.3. Properties of HiPIMS-deposited Al$_{80}$Sc$_{20}$N thin films

### 2.3.1 Morphology and abnormally oriented grains

Given that the solubility limits for all examined samples exceeded 20%, we decided to deposit AlScN films with a homogenous Sc concentration of 20% for all promising deposition schemes for further investigation. At first, we investigated the microstructures of films deposited with Al-HiPIMS/Sc-DCMS with grounded and synchronized substrate biases. **Figure 4** shows AFM images for films deposited without substrate rotation (a-b), with substrate rotation (c-d), and with substrate rotation on a Ti/Pt bottom layer (e-f). The triangular-appearing structure protruding out from the surface in the images are abnormally oriented grains (AOGs), which diverge from the otherwise uniform texture. AOG are typically formed due to c-axis instability occurring during growth in which the cubic ScN nuclei get incorporated in the (002) AlScN wurtzite crystal during growth. Over time, Al gets incorporated in the c-ScN crystallites, prompting it to revert to the wurtzite phase. This transformation results in the loss of original orientation and the formation of tilted AOGs.[8] The AOGs are not aligned with the rest of the grains and thus lower the overall piezoelectric performance of the film. With increasing thickness, these grains enlarge and take up large fractions of the film volume. Therefore, it is important to address the formation of these grains and ways to prevent their nucleation. The AOGs are more prominent in films deposited without rotation, contrary to other films. This is primarily because of the shadowing effect caused by protruding AOGs, which leads to further accumulation of Sc at the protruding grain boundaries. Strikingly, the AOG can be completely obliterated with the application of substrate biasing (sync or DC), due to increased ad-atom mobility and compressive stress in the film.[46] High ad atom mobility allows the atom to arrange in low-energy basal planes avoiding the nucleation of disoriented grains in other directions. On the other hand, compressive stress in the film suppresses the growth of AOG by densifying the film. Both factors have been shown to be effective in the removal of these grains, however, it is difficult to pinpoint which factor is more prominent here. Previous studies have demonstrated that the bottom Pt layer also plays a role. Here epitaxial matching of AlScN(002) and Pt(111) occurs when the w-AlScN crystal is rotated 30° along the c-axis relative to the Pt(111) template.[47] Thus, in order to facilitate piezoelectric measurements and enhanced nucleation, the isolated films were grown on a 100-nm-thick Pt layer for further investigations (see **Figure 1**). With Pt as a bottom layer, the enhanced nucleation can be seen with the enlargement of grains of the film. The pronounced nucleation with the help of the Pt layer in combination with substrate rotation and biasing can help in obtaining crystallized films with no AOGs.

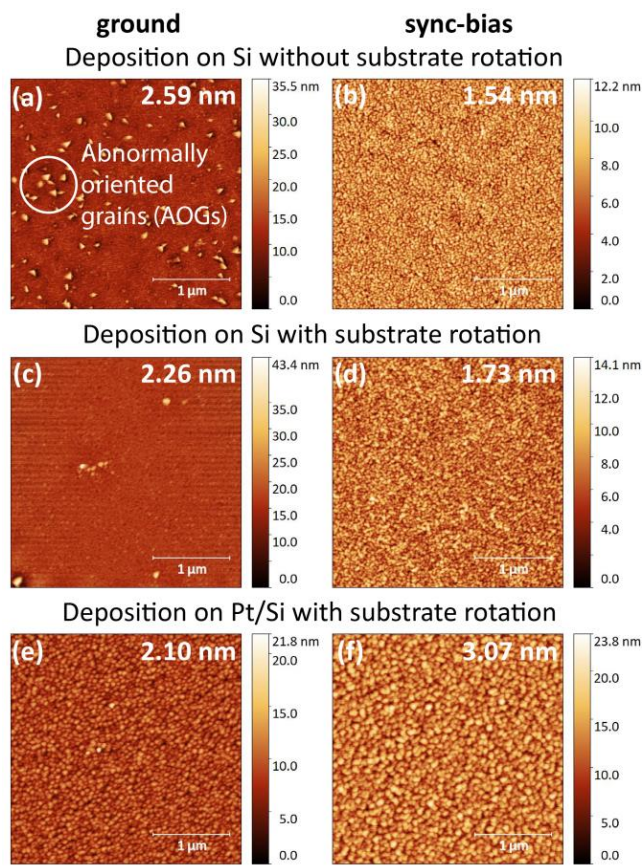

**Figure 4.** AFM imaging of Al$_{0.8}$Sc$_{0.2}$N films deposited with the grounded and sync-bias approach for substrate rotated and still during the deposition. The images show changes in the formation of abnormally oriented grains as well as the RMS surface roughness. The shadowing effect, prominent when the substrate is stationary, increases Sc accumulation over grains. Rotating the substrate during deposition helps mitigate this effect. In addition, the high ad atom mobility of atoms and compressive stress induced by ion bombardment also suppress the growth of AOGs.

### 2.3.2 Residual Stress and Texture

Following the combinatorial screening, we proceeded to investigate isolated films with a homogeneous 20% Sc concentration concerning their residual stress and out-of-plane texture. The films investigated here showed no detectable oxygen contamination even after exposure to air due to their compact microstructure (see Supporting Information S3). The data for all samples, along with the solubility limits estimated from combinatorial screening, are plotted in **Figure 5(a)** and arranged in order of stress from tensile to compressive (left to right). All the films were completely textured and highly crystalline, which was verified from the XRD patterns and pole figures attached in Supporting information S2. For comparison, data for AlN films deposited using HiPIMS with grounded and synchronized biasing are also discussed. Strikingly the residual stress changes significantly for different deposition schemes, which result in more or less energetic ion bombardment. This highlights the ability to tailor stress in the films by adjusting the HiPIMS deposition parameters and the respective ion acceleration. The most pronounced out-of-plane texture is observed for films with no or slight tensile stress. The FWHM of the rocking curve evidently increases with the residual stress in the film, regardless of whether it is tensile or compressive. This is attributed to the dislocations, grain elongation, and microstructural defects caused by increased stress in the system.

In addition to changes in the out-of-plane texture, we observe a strong correlation of the films' stress with the estimated solubility limits from the combinatorial screening of the libraries of respective deposition schemes. In **Figure 5(a)** and **(b)**, the solubility limit from combinatorial screening can be seen to vary linearly with the residual stress of the film. The effect of stress on the maximum solubility of cubic systems in wurtzite has also been studied earlier based on ab initio thermodynamic and kinetic models for other material systems such as AlTiN, AlVN, and AlCrN.[48], [49] Residual stress, deposition rates, thermodynamic and kinetic factors have been found to be the driving factors for the variation in solubility observed in these systems. Greczynski *et al.* reported the variance of maximum solubility by subplantation of energetic metal ions onto the high mobility surface zone of the growing film.[50] However, due to the low substrate bias potentials chosen in this investigation, particularly to avoid the formation of defects and large compressive stresses, the effect of ion subplantation is not dominant and can be ruled out. Deposition rate and microstructural effects can also play an important role in determining the maximum solubility of the system. All investigated films exhibit similar compact columnar microstructure and deposition rates, except the film deposited in purely DCMS mode, which tends to have a columnar structure with open grain boundaries.[28] The presence of high compressive stress leads to a densification of the film, thereby favoring the formation of a more compact cubic lattice. Thus, it leads to reduced maximum solubility of Sc lattice in wurtzite AlN, i.e. precipitation of c-ScN at lower Sc concentrations. From our investigation we conclude, that for films exhibiting compact microstructures, intrinsic stress is the main driving force for the variation of the solubility limits.

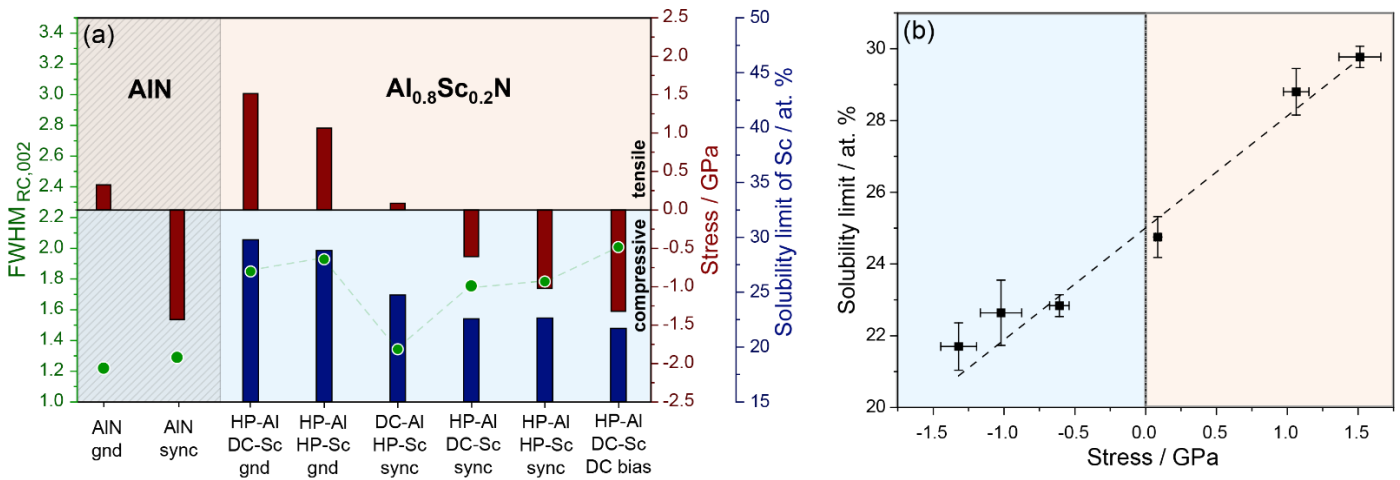

**Figure 5.** (a) Plot summarizing the FWHM of rocking curve, residual stress and solubility limit estimated from combinatorial screening for AlN and AlScN films deposited with different combination of power supplies and substrate biasing. Here HP and DC are used as abbreviations for sputter guns operated in HiPIMS and DCMS mode, respectively. (b) Change in solubility limit with stress obtained for films discussed in Figure 5(a).

### 2.3.3 Piezoelectric response

Subsequently, we measured the piezoelectric response of the above-discussed films and determined the clamped piezoelectric coefficients, $d_{33,f}$. The measurements were performed using double-beam laser interferometry (DBLI) at aixACCT GmbH. The reported values measured with this technique tend to be slightly lower than those measured with mechanical methods (e.g. Berlin court method). [51] DBLI is considered one of the most accurate characterization techniques for piezoelectric thin films as it fully accounts for the deformation of the substrate during actuation.[52]

The measured piezoelectric coefficients for the HiPIMS deposited AlScN are plotted in **Figure 6** along with comparable DBLI measured experimental values obtained from the literature and DFT predictions from Caro. et. al.[53]–[55] In addition, we report $d_{33,f}$ values for AlN films.[28] To our knowledge, these are the first reported piezoelectric coefficients for AlN and AlScN films deposited with HiPIMS processes. The $d_{33,f}$ of AlN films deposited with HiPIMS were found to vary from 3.59 to 3.92 pm/V, while the $d_{33,f}$ of AlScN films varied from 6.31 to 8.81 pm/V. Strikingly, despite the shallow sputter angles and high working distances the response of our HiPIMS deposited films is on par with experimental values by Mertin *et al.* and Tsubouchi *et al.*[53], [54] This is particularly interesting as these reference values were achieved using co-planar deposition geometry (i.e. on-axis sputtering at low working distances) in an industrial sputter coater, closely resembling the current state-of-the-art in the industry.

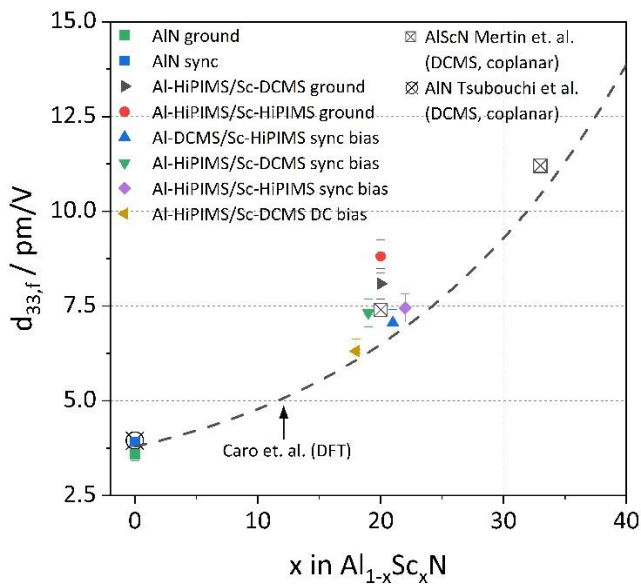

**Figure 6:** Piezoelectric coefficient, $d_{33,f}$ for investigated $Al_{0.8}Sc_{0.2}N$ samples along with DFT predictions and experimental values reported in the literature. The $d_{33,f}$ values closely match with the DFT predictions [55] and experimental values of samples deposited in coplanar geometry [53], [54].

Some minor differences in the response are observed among the HiPIMS-deposited films. It is interesting to note that the piezoelectric coefficients of the films are not overly sensitive to residual stress. Since all the investigated films were relatively well oriented in the c-axis, the effect of crystallinity on piezoelectric coefficients is not pronounced. On a closer look, the $d_{33,f}$ of all metal-ion-synchronized films are minutely lower than the ones with the grounded substrate, while the one with DC-bias is even lower than the sync bias samples. The effect is very little, presumably due to the gentle conditions used for the deposition of these functional films. It is speculated that the minute decrease in the coefficients could be due to the formation of point defects in the films due to substrate biasing and compressive stress. This effect of stress on the piezoelectric response has also been observed by researchers earlier.[56], [57] On the other hand, the use of synchronized biasing offers the advantages of tuning the stress state of the film along with its application to coat structured surfaces or deposition of these functional films in oblique conditions. The ability to deposit high-quality piezoelectric thin films at shallow deposition angles and moderate temperatures demonstrates the great potential of ionized PVD processes for functional defect-sensitive coatings.

### 2.3.4 Deposition on structured surfaces

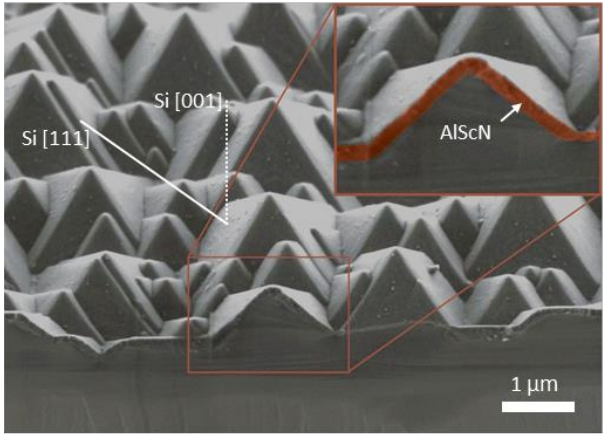
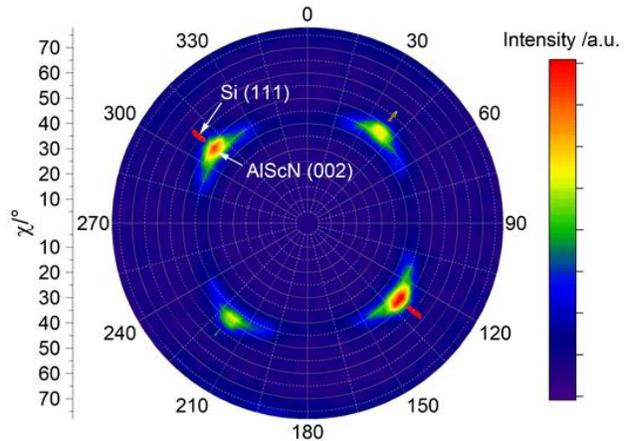

**Figure 7**: (a) SEM cross-section of AlScN films grown on structured Si wafer. The wafer has micro-sized pyramids with [111] facets formed by chemical etching of a [001] wafer. The cross-section reveals a uniform film thickness. (b) XRD pole figure measurement on AlScN (002) confirms that the films grow with a pronounced c-axis texture on the [111] facets of the pyramids confirming that the films' texture is maintained even at shallow deposition angles.

The HiPIMS technique, combined with a sync substrate biasing approach, offers a convenient method for depositing films on complex structures by deflecting ions towards the substrate and increasing the adatom mobility. While this technique has long been utilized by the industry for depositing hard coatings in micro-trenches and 3D structures, its application in growing crystalline functional coatings has not been explored yet.[58], [59] Thus, to verify the conformal growth of crystalline AlScN films over structured surfaces, we deposited a 150-nm-thick film on an etched Si wafer with pyramids featuring facets in the [111] direction. The Al target was positioned under the substrate and operated in HiPIMS mode, while Sc was deposited using two sputter guns at an oblique angle operating in DCMS mode (see Supporting information S4). A synchronized substrate bias potential was applied. The substrate was rotated throughout the deposition process to ensure uniform Sc alloying and conformal deposition. This setup was intentionally chosen to ensure the continuous arrival of Al and Sc flux at the substrate, mitigating the potential formation of stacked AlN and ScN layers due to shadowing effects that could occur if both Al and Sc were operated at oblique angles.

The SEM cross-sections in **Figure 7(a)** illustrate the uniform growth of the film, with the pyramids maintaining their shape even after film deposition, indicating consistent film growth over the facets. To assess the texture of the films over the pyramids, pole figure measurements were conducted on the bare Si substrate and on the coated substrate, as shown in **Figure 7(b)**. The Si(111) pole figure shows sharp peaks at 4 points corresponding to the 4 facets of the pyramids, and similarly, the AlScN film grown over these facets also shows corresponding 4 points on the pole figure indicating growth of c-axis oriented films on all 4 facets. Only a slight difference of about 5° in χ between the Si(111) and AlScN(002) suggests a minor misalignment of grains relative to the facets' normal. In the future, this misalignment could be further reduced by sputtering from a compound target in coplanar geometry and further optimization of the substrate bias potential pulse pattern.

## 3. Conclusion

The demand for MEMS thin films with tailored functionality is increasing rapidly. Future applications will include increasingly complex sample geometries and constraints for deposition temperatures. It has recently been shown that metal-ion synchronized HiPIMS can offer significant advantages for the deposition of highly textured thin films at moderate temperatures, even at shallow deposition angles or on structured surfaces.[28] In this study, we investigated the feasibility of this concept for the reactive co-sputtering of AlScN and particularly for the fabrication of piezoelectric devices. To this end, different combinations of sputtering modes (DCMS and/or HiPIMS) along with varying substrate biasing approaches (ground, DC bias, sync-bias) were tested. For a rapid screening of the synthesis parameter space, a combinatorial synthesis with automated characterization and data analysis was employed. The combinatorial screening helped assess the relative crystallinity of films and of solubility limit of Sc for a number of different synthesis approaches. This screening showed a significant improvement in crystallinity for HiPIMS approaches compared to conventional sputtering, at the cost of Sc solubility. Based on this screening, single-phase $Al_{0.8}Sc_{0.2}N$ films were deposited on smaller Si substrates with homogenous composition for promising conditions. It was found that the growth of AOGs can be suppressed by substrate rotation and application of negative biasing on the substrate. Detailed analysis of the structural properties revealed, that the stress state of the films can be tuned over several GPa using a combination HiPIMS deposition and substrate biasing. Strikingly, a direct correlation between the maximum solubility of Sc and the residual stress in the films was established. Here, we find that the compressive stress leads to the densification of the film and a change in the c/a ratio in the hexagonal lattice of w-AlScN, ultimately resulting in the precipitation of the denser c-ScN phase. Finally, the piezoelectric response was investigated on thin-film device structures using DBLI. Despite the non-optimized chamber geometry (i.e. shallow deposition angles and high working distances) the piezoelectric coefficient $d_{33,f}$ were found to be on par with the current state-of-the-art achieved in production systems. The $d_{33,f}$ show similar values for all HiPIMS approaches. Although all the films exhibited similar piezoelectric responses varying from 6.3-8.8 pm/V, the films deposited with sync-bias provide us with an opportunity to tune the stress of the system, deposit films with compact microstructure and provide the ability to coat on structured substrates. The conformal coverage of AlScN films over a structured Si wafer was further confirmed with the help of SEM imaging and pole figure measurements confirmed c-axis oriented growth on the [111] facets of the pyramid structure. Overall, the results of this study demonstrate the potential of advanced ionized PVD approaches like MIS-HiPIMS for next-generation MEMS technologies. Implementing such deposition techniques might enable exciting new applications for piezoelectric and ferroelectric thin films in the future, in particular for applications requiring low substrate temperatures or devices with complex geometries.

## 4. Experimental

AlScN films were deposited using a custom-built sputter chamber, AJA International, ATC-1800, with a base pressure of < $10^{-6}$ Pa. The films were deposited on p-type Si (100) wafers with unbalanced magnetrons equipped with 2 inch Al (Lesker, purity: 99.999 at.%) and Sc target (Plasmaterials, purity: 99.9 at.% & Hunan Advanced Metal Material Co. Ltd, purity: 99.9 at.%, O < 800 ppm). The magnetrons were aligned with a sputter angle of 26° with respect to the substrate normal and the working distance was set at 12 cm. A closed-field magnetic configuration was used between the opposite magnetrons to reduce the amount of plasma heating.[28], [60], [61] Nitrogen was routed directly to the targets to ensure easier poisoning of the targets while the Ar gas was supplied away from the target in the chamber. The flow rates of Ar/$N_2$ and working pressure is maintained constant at 20 sccm and 12 sccm, respectively. The depositions were performed at 300°C and at a working pressure of 3.5 Pa. The substrate holder is heated from the back using 5 halogen lamps, resulting in homogenous heat distribution across the holder. The substrate is maintained at a constant temperature of 300°C. Before the deposition the Si substrates are etched by RF Ar plasma at 30 W for 8 min to remove the native oxide layer. The power and current densities for DCMS and HiPIMS for each sputter gun are summarized in Table 1. The substrate holder is either grounded or negatively biased depending on different deposition approaches, which are described further in this section.

**Table 1 :** Power and current densities for Al and Sc sputter guns operated in DCMS and HiPIMS mode

|  | Power density (W/cm$^2$) | Current density (mA/cm$^2$) |
|---|---|---|
| **Al-DCMS** | 6 | 23-26 |
| **Al-HiPIMS** | 5 | 450-550 |
| **Sc-DCMS** | 0.85 | 3.5-5.5 |
| **Sc-HiPIMS** | ~1.5 | 100-150 |

For the synchronization of the substrate bias, the time-of-flight of the Al and Sc ions was first estimated by time-resolved mass spectrometry measurement using a Hiden Analytical EQP-300. The orifice of the spectrometer, 50 µm in diameter, was grounded and placed at the working distance while facing the sputter gun. The triggering signal was provided by the pulsing unit of the HiPIMS power supply attached to the target. The gate width was set to 5 µs, consistent with the step size of measurement. The time-resolved measurements were performed for each ion species, including $^{14}$N, $^{36}$Ar, $^{27}$Al and $^{45}$Sc. A less abundant isotope of Ar is used here to avoid the saturation of the detector. The time-of-flight in the mass spectrometer is calibrated by applying a gating potential at the driven front end of the spectrometer. A more detailed description of this method can be found in our previous publications.[28]

For the films' deposition, the Al and Sc sputter guns were operated in multiple combinations of sputtering modes, namely HiPIMS-AlN/DCMS-ScN, DCMS-AlN/HiPIMS-ScN, DCMS-AlN/DCMS-ScN, HiPIMS-AlN/HiPIMS-ScN. DCMS is carried out using a 750 W DC power supply by AJA International (DCXS 750). HiPIMS is carried out using Ionautics pulsing units and power supplies (HiPSTER 1 bipolar). The 10 µs HiPIMS pulses with a frequency of 7.5 kHz on the targets were synchronized to start at the same time using an Ionautics synchronization unit.

For the substrate, different biasing strategies are evaluated. These included a grounded substrate, a constant bias of -30 V (DC-bias) or a synchronized bias pulse of -30 V. In the latter mode, the substrate pulse was synchronized to the metal-rich part of the plasma. Specifically, a 40 µs substrate bias pulse of -30 V is applied with an offset of 20 µs using the same bipolar HiPIMS power supply and is actively regulated to 0 V between the pulses. The synchronization scheme with the I-V curve of the HiPIMS discharges is shown in **Figure 2**.

For the combinatorial depositions, the films were deposited without substrate rotation on a 2×2 in Si (100) wafer, thus obtaining compositional gradient on one axis. The wafer was clamped uniformly to the holder with additional masks to homogenize the temperature throughout the wafer. Each combinatorial library has 3 rows with each row consisting of 9 samples with increasing composition of Sc in one direction, thus giving 27 samples in total for one deposition, as illustrated in **Figure 1**. The data from multiple rows are investigated in order to rule out geometric effects from localized plasma heating. Based on the results from the combinatorial studies, the promising set of conditions were chosen for deposition of isolated samples on a smaller 13×13 mm wafer. The isolated samples were deposited for a fixed Sc composition of ~ 20±2 at.% with substrate rotated at 40 rpm. The stack of layers is illustrated in **Figure 1** for these samples. A 10 nm Ti layer act as an adhesion layer and also prevents the formation of PtSi at interface. The Pt layer over it acts as a bottom electrode for piezoelectric measurements and also enhances the nucleation due to the epitaxial relationship and reduced lattice mismatch in comparison to Si lattice.[47] Lastly, the top electrode pads were deposited over the AlScN and AlN films with the help of a shadow mask with square pad areas of 16, 36 and 64 µm$^2$. The films used for the fabrication of piezoelectric devices showed no detectable oxygen contamination (measured via X-ray photoelectron spectroscopy (XPS) depth profiles) even after exposure to air due to their compact microstructure. The AlScN films deposited with a lower purity Sc target exhibited approximately 2-3% bulk oxygen, resulting in a slightly reduced piezoelectric coefficient. These films are not further discussed in the main text and the $d_{33,f}$ comparison along with XPS depth profiles of these films can be found in Supporting information S3.

X-ray diffraction (XRD) analysis of the films was performed using a Bruker D8 in Bragg-Brentano geometry and Cu-k$\alpha$ radiation. The estimation of crystalline quality and solubility limit was done for the combinatorial libraries using automated XRD θ-2θ measurements. The composition of the combinatorial libraries and films were estimated using automated X-ray fluorescence (XRF) measurements. The XRF data for chemical composition of films was calibrated based on the composition obtained through X-ray photoelectron spectroscopy (XPS, PHI-Quantera, Al k$\alpha$ radiation) for a few combinatorial libraries transferred in UHV conditions directly after the deposition to the XPS. The large datasets were then analyzed using CombIgor data analysis package in Igor pro software. The morphology and disoriented grains present in the films were visualized using Bruker nanoscope atomic force microscopy (AFM). The measurement were done in ScanAsyst mode with silicon cantilever tips. $3 \times 3$ μm$^2$ AFM images with a pixel resolution of $512 \times 512$ were captured using Nanoscope software and later processed in Gwyddion. The thickness of the films were measured using Dektak profilometer. The residual stress in the films were calculated using the wafer curvature method, by measuring the peak position of Si (100) peak via rocking curve measurements at several positions on the wafer. Stoney's equation was used further to calculate the residual stress in the films. The equation and coefficients used for the calculation can be found in Supporting information S5.

The piezoelectric response was measured using double-beam laser interferometry (DBLI) at aixACCT GmbH. This method uses laser interferometry to measure the surface displacement of the sample relative to the applied electrical voltage. During the measurement, the bending of the substrate is taken fully into consideration by using a double-beam configuration, which is often overlooked by piezometers employing the Berlin-court method, resulting in the measurement of inaccurate piezoelectric values. The measurements were performed on different electrode sizes (i.e. $400 \times 400$ μm$^2$, $600 \times 600$ μm$^2$ and $800 \times 800$ μm$^2$) and the effective clamped piezoelectric response $d_{33,f}$ was calculated based on the procedure described by Sivaramakrishnan *et al*.[62]

The microstructural characterization of AlScN films on structured silicon wafers was performed using a Zeiss Gemini 460 Scanning Electron Microscope (SEM), equipped with a secondary electron detector (SE detector) operated at an accelerating voltage of 2 kV and a sample current of 50 pA. The samples were cleaved in air and secured onto a SEM sample holder using carbon tape.

# Acknowledgements
J. P. acknowledges funding by the SNSF (project no. 200021_196980). Bertrand Paviet-Salomon is gratefully acknowledged for providing the structured Si-substrates. The authors also acknowledge the help from Sebastian Bette with the DBLI measurements.

# Author contributions
**J. P.:** Conceptualization, Investigation, Methodology, Formal analysis, Visualization, Writing - Original Draft; **K.T.:** Investigation, Writing – Review & Editing; **T.S.K.:** Investigation, Writing – Review & Editing **R.K.:** Investigation, Writing

– Review & Editing **S.S.:** Conceptualization, Supervision, Methodology, Formal analysis, Visualization, Funding acquisition, Writing – Review & Editing

**Supporting information for**

# Deposition of highly-crystalline AlScN thin films using synchronized HiPIMS – from combinatorial screening to piezoelectric devices

Jyotish Patidar[1], Kerstin Thorwarth[1], Thorsten Schmitz-Kempen[2], Roland Kessels[2], Sebastian Siol[1*]

[1] Empa, Swiss Federal Laboratories for Materials Science and Technology, Dübendorf, Switzerland

[2] aixACCT Systems GmbH, Aachen, Germany

*Corresponding author:*

*Sebastian Siol, Sebastian.Siol@empa.ch*

# S1: Estimation of solubility limits using the disappearing phase method

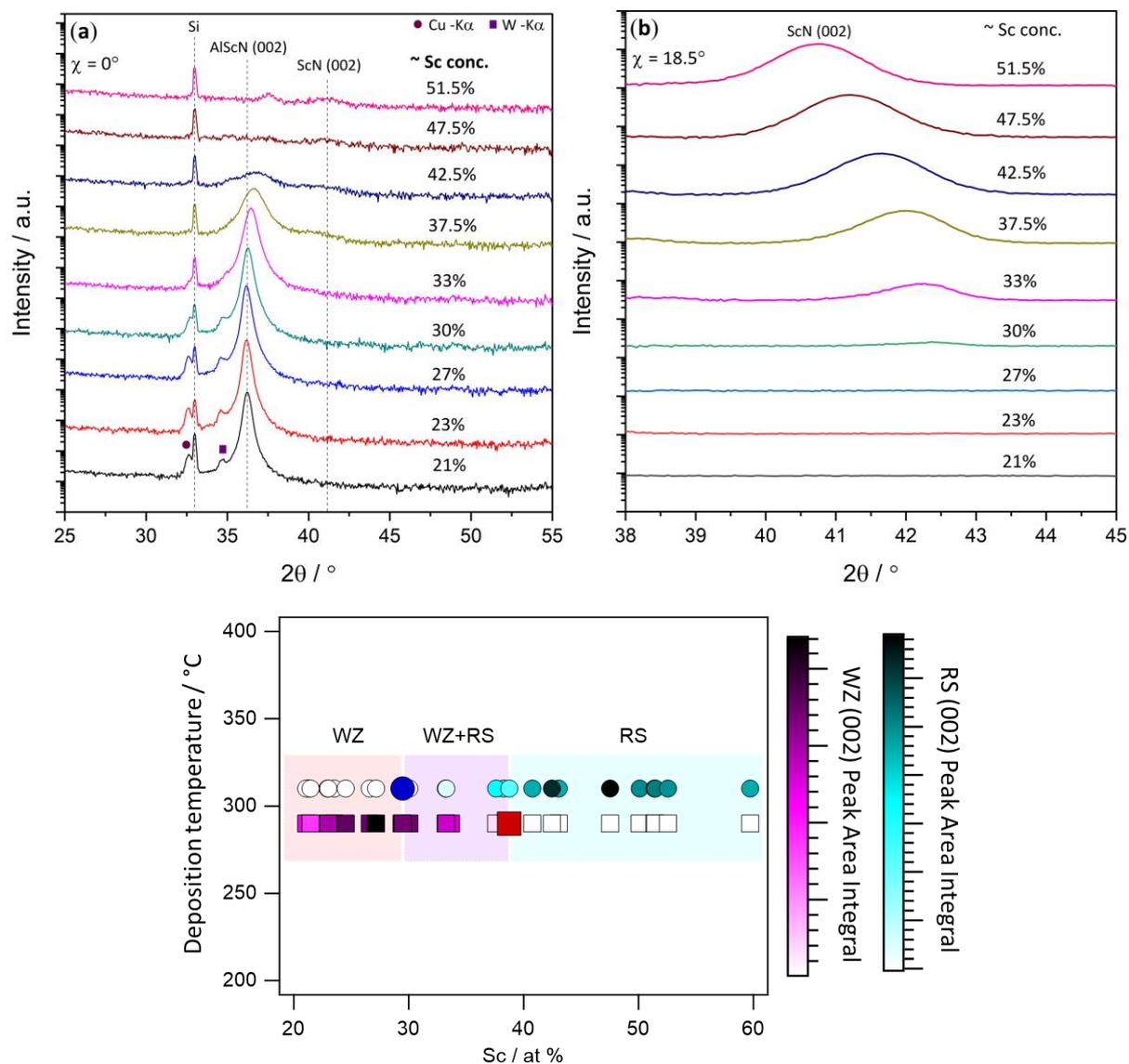

**Fig. S1**: XRD patterns of AlScN films measured for different Sc compositions in a combinatorial library **(Al-HiPIMS/Sc-DCMS (ground))** for (a) χ = 0° and (b) χ = 18.5°. Here χ is the tilt of the sample in XRD equipment corresponding to the tilt of the grains of the ScN deposited in DCMS mode. (c) Based on the XRD screening, the fit integral of AlScN (Wurtzite, WZ) and ScN (Rock-salt, RS) XRD peaks are taken and plotted to estimate the solubility limit for a particular deposition condition. The red square marks the solubility limit for this library.

## S2: XRD analysis of Al$_{0.8}$Sc$_{0.2}$N films

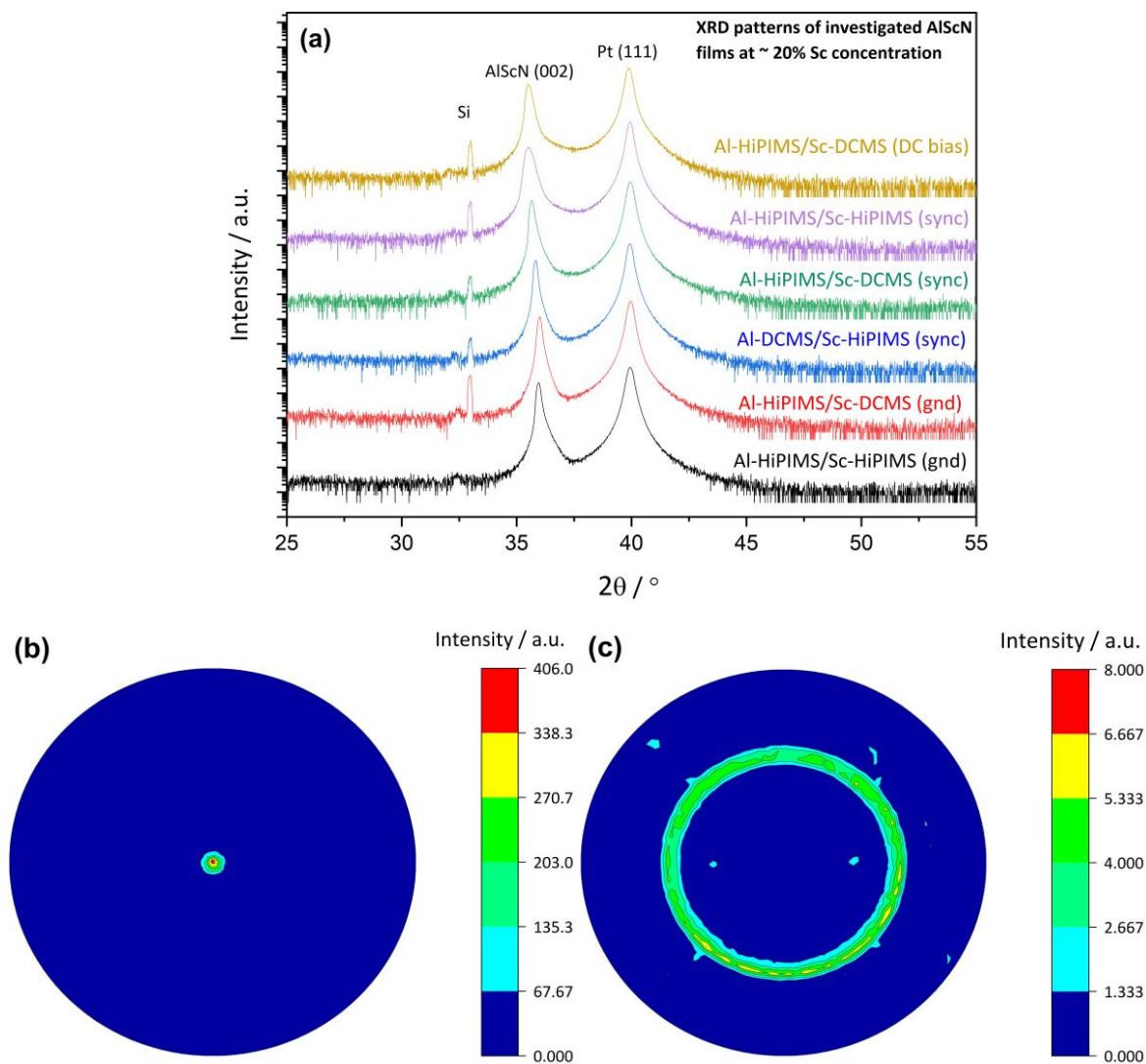

**Fig. S2**: (a) XRD patterns of investigated AlScN films deposited with different power supplies and substrate biasing strategies. (b-c) Pole figures at (002) and (103) peak of AlScN film deposited with Al-HiPIMS/Sc-DCMS (sync bias). The figures confirm the (002) out-of-plane texture and random in-plane orientation of the film. All the investigated films in this study are fully textured with clear (002) out-of-plane orientation.

# S3: Comparison of samples with varying oxygen content

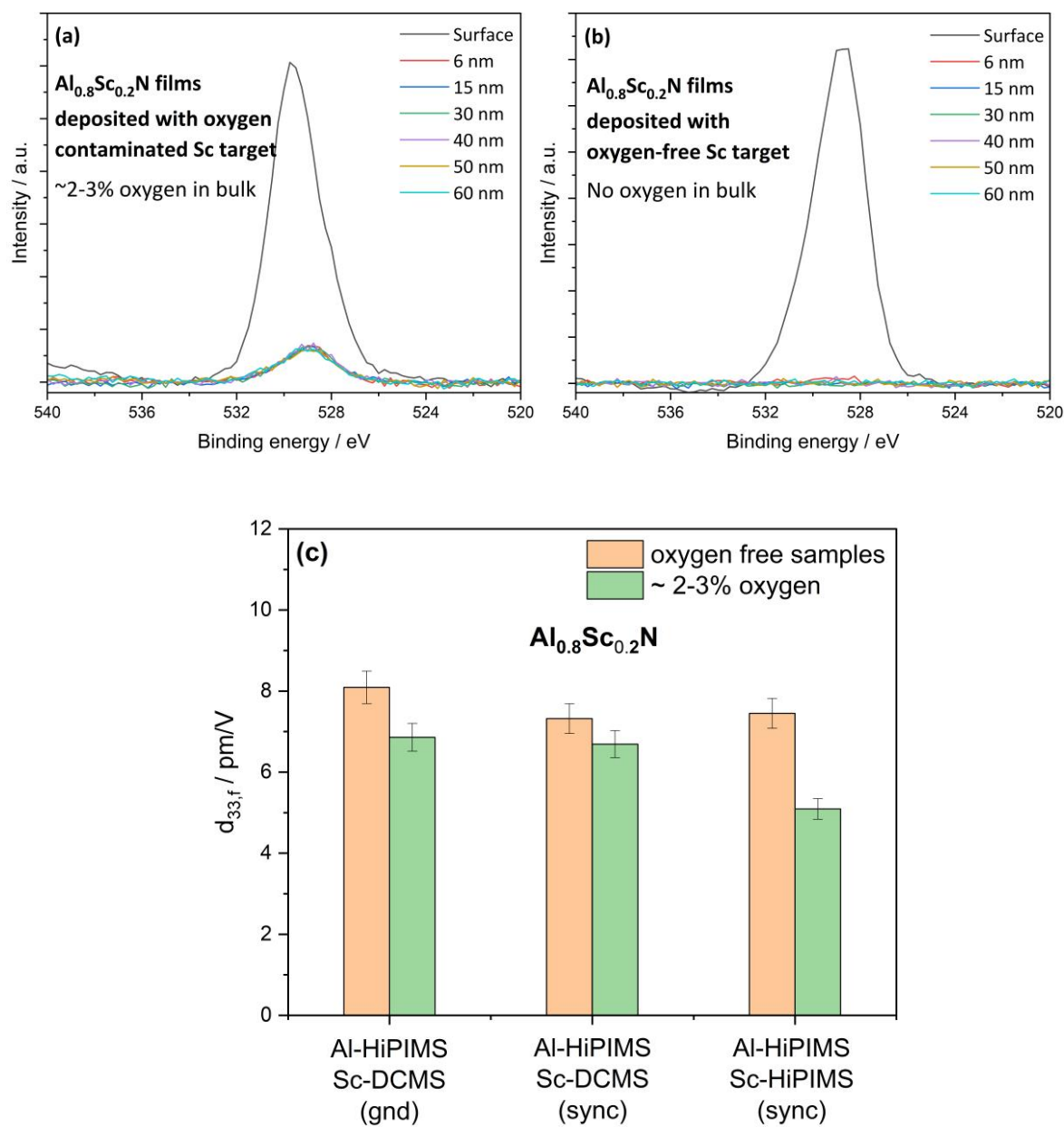

**Fig. S3**: The AlScN films with 20% Sc were also deposited with an oxygen-contaminated Scandium sputter target. The films deposited using this target showed oxygen levels of about 2-3% in the bulk. This was confirmed by XPS depth profiles. Figures (a) and (b) show the O 1s core level emission for films deposited with clean and contaminated sputter targets, respectively. A slightly reduced piezoelectric response was measured for AlScN films with oxygen in bulk, which is attributed to the decreased crystalline quality and change in texture during the film growth. [1,2]

## S4: Sputter gun geometry while deposition on structured substrate

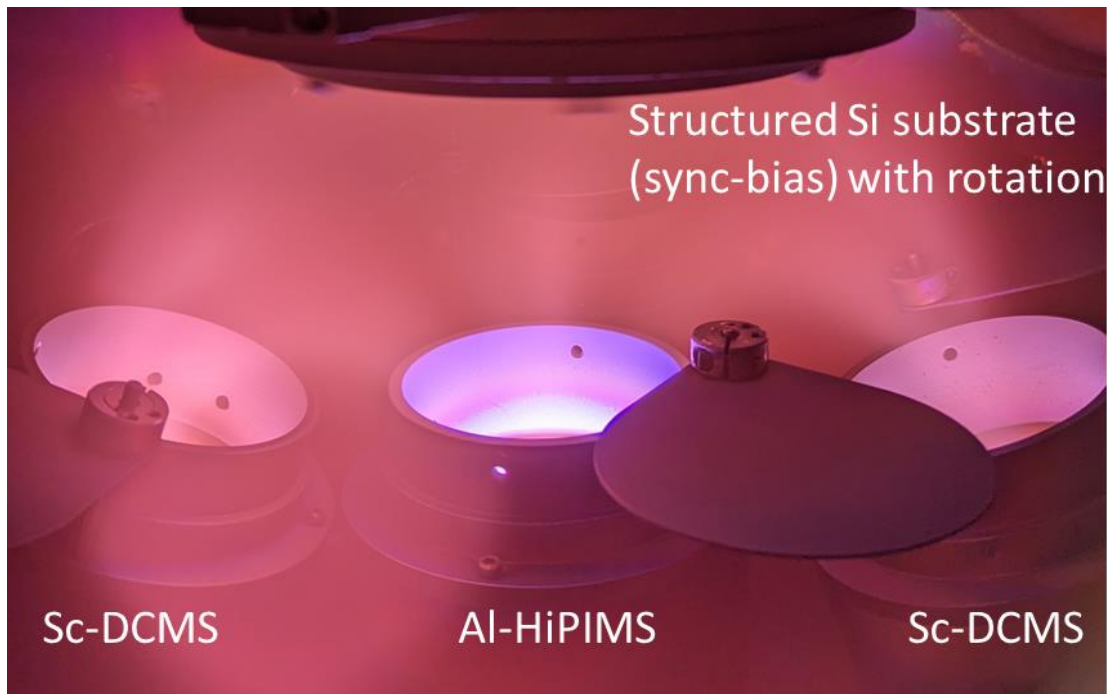

**Fig. S5**: Image of actual deposition of AlScN film on the structured wafer. Here the middle sputter gun is Al, which is operated in HiPIMS mode in coplanar geometry while the two Sc guns on the side are operated in DCMS mode with oblique-angle deposition. The substrate is rotated at 40 rpm throughout the deposition to ensure uniform flux of both Al and Sc.

# S5: Residual stress measurement through XRD by estimating wafer curvature

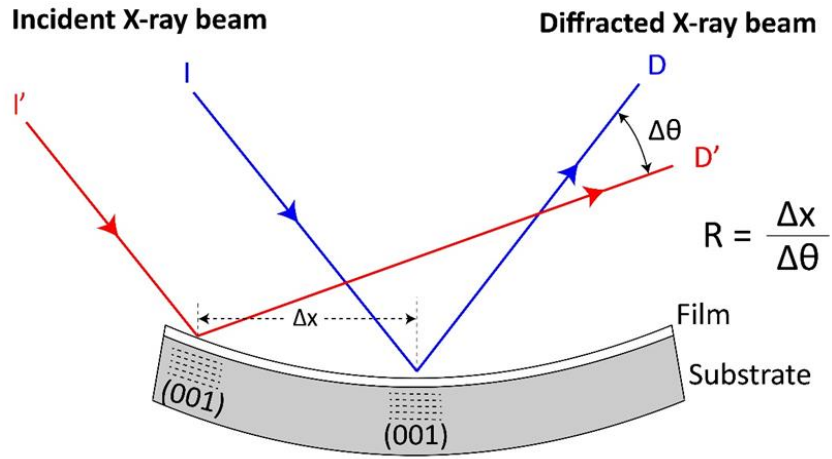

**Fig. S4**: Schematic of measurement for wafer curvature using XRD.

The residual stress in the films was measured by estimating the curvature of the wafer after and before the deposition. At first, the Si (002) peak position was determined via rocking curve measurements in XRD on several positions on the wafer. The substrate curvature (R) was determined by calculating the slope of the plot of measurement positions (x) against the peak position of the rocking curve (θ). The curvature was then translated to stress (σ) with the help of Stoney's equation,

$$\sigma = \frac{1}{6} \frac{E_s}{1-\nu_s} \frac{t_s^2}{t_f} \left( \frac{1}{R} - \frac{1}{R_o} \right)$$

Where $E_s$ and $\nu_s$ are the elastic modulus and Poisson's ratio for the substrate. For Si(001), the value of ($E_s / 1 - \nu_s$) is 181 GPa at room temperature.[3] $t_s$ (≈ 0.540 mm) and $t_f$ are the thickness of the substrate and film, and $R$ and $R_o$ are the curvature of the wafer before and after the deposition.